# A Molecular Mechanics Study of Morphologic Interaction between Graphene and Si Nanowires on a SiO$_2$ Substrate

Zhao Zhang[a], Teng Li[a,b,*]

[a]Department of Mechanical Engineering, University of Maryland, College Park, MD 20742

[b]Maryland NanoCenter, University of Maryland, College Park, MD 20742

## Abstract

In this paper, we study the morphologic interaction between graphene and Si nanowires on a SiO$_2$ substrate, using molecular mechanics simulations. Two cases are considered: 1) a graphene nanoribbon intercalated by a single Si nanowire on a SiO$_2$ substrate and 2) a blanket graphene flake intercalated by an array of Si nanowires evenly patterned in parallel on a SiO$_2$ substrate. Various graphene morphologies emerge from the simulation results of these two cases, which are shown to depend on both geometric parameters (e.g., graphene nanoribbon width, nanowire diameter, and nanowire spacing) and material properties (e.g., graphene-nanowire and graphene-substrate bonding strength). While the quantitative results at the atomistic resolution in this study can be further used to determine the change of electronic properties of graphene under morphologic regulation, the qualitative understandings from this study can be extended to help exploring graphene morphology in other material systems.

---

[*] Author to whom correspondence should be addressed. Electronic mail: LiT@umd.edu

## 1. INTRODUCTION

The experimental discovery of graphene[1] has inspired a surge of interest in developing its application toward novel nanoelectronic devices[2-7]. The atomically thin structure of graphene dictates the strong correlation between the electronic properties and the morphology of graphene[8-11]. For example, bending-induced local curvature in graphene changes the interatomic distances and angles between chemical bonds, and thus leads to changes in the graphene's band structure[9]. Therefore, the intrinsic random corrugations in freestanding graphene[12] may result in unpredictable fluctuation of electronic properties that is undesirable for graphene-based nanodevice applications. Recent experiments[13-17] and simulations[18-20] reveal that the morphology of graphene can be controlled via extrinsic regulation, which envisions a promising pathway toward tunable electronic properties of graphene. In this paper, using molecular mechanics simulations, we study the morphology of graphene nanoribbons and flakes regulated by Si nanowires on a $SiO_2$ substrate. The results from the present study can shed light on the structural design of graphene-based field effect transistors and other nanoelectronics components.

Recent experiments[13, 14] show that the interaction between graphene and its underlying substrate can effectively suppress the intrinsic random ripples in the graphene. As a result, the graphene can partially conform to the substrate surface corrugations. Inspired by these experimental observations, recent modeling and simulations [18-21] have predicted the graphene morphology regulated by various substrates with engineered nano-scale surface features. The substrate-regulated morphology of graphene is governed by the interplay between the graphene-substrate interaction energy and the deformation energy of the graphene due to bending and stretching. Emerging from these theoretical studies is an effective approach to controlling the graphene morphology and thus its electronic properties via external regulation.

Nanowires (e.g., Si) of diameters as small as 1 nm have been successfully fabricated[22], which can be used as building blocks of nano-electronic devices. Furthermore, ultrathin nanowires patterned on a substrate surface[23] can potentially serve as scaffolds to regulate the graphene morphology at an even higher resolution that approaches the atomic feature size of graphene (e.g., the carbon-carbon bond length in graphene is about 0.14 nm). To this end, we recently extended our previous continuum mechanics energetic framework[19] to quantitatively determine the graphene morphology regulated by nanowires patterned in parallel on a substrate surface[20]. In the extended modeling framework, the graphene-nanowire interaction energy is incorporated into the energetic interplay. This extended continuum mechanics model can shed light on predicting the graphene morphology modulated by nanowires of diameters comparable to or modestly smaller than the spatial resolution of engineered substrate surfaces (e.g., as low as about 10 nm). It is expected that, however, the continuum mechanics model cannot capture the full characteristics of the graphene morphology regulated by ultrathin nanowires (e.g., with diameters on the order of 1 nm). Furthermore, the continuum model cannot determine the exact positions of each carbon atom in graphene at the equilibrium morphology. Such information will be needed in further first principle calculation of the electronic properties of the graphene.

To address the above concerns, in this paper, we carry out molecular mechanics (MM) simulations to study the morphological interaction between graphene nanoribbons/flakes and Si nanowires on a SiO$_2$ substrate. The rest of the paper is organized as follows. Section 2 describes the computational model; In Section 3, we consider two simulation cases: 1) interaction between a graphene nanoribbon and a Si nanowire on a SiO$_2$ substrate; 2) interaction between a blanket graphene flake and an array of Si nanowires patterned in parallel on a SiO$_2$ substrate; Discussions and summary are given in Section 4.

## 2. COMPUTATIONAL MODEL

The nanowire-regulated morphology of graphene on a substrate is governed by the interplay among the following energies: the strain energy of the graphene due to bending and stretching, the graphene-nanowire interaction energy, and the graphene-substrate interaction energy. Both non-bonding interaction energies can be characterized by van der Waals force, which minimizes at a certain equilibrium distance between the graphene and the nanowires (or the substrate). On the other hand, the graphene strain energy minimizes when the graphene remains flat and increases monotonically as the graphene corrugates to conform to the nanowires and the substrate. As a result, the total energy of the graphene-nanowire-substrate system minimizes when the corrugated graphene reaches its equilibrium morphology.

Figure 1 depicts the two configurations simulated in this paper: (1) a graphene nanoribbon of finite width in *y*-direction on a SiO$_2$ substrate with a Si nanowire intercalating in between (Fig. 1a) and (2) a blanket graphene flake intercalated by an array of Si nanowires patterned in parallel on a SiO$_2$ substrate (Fig. 1b). Given the periodicity of these two configurations, only the portion of the graphene marked by dash lines and the corresponding nanowire and substrate underneath are simulated. In the MM simulations, periodic boundary conditions are applied to the two end surfaces in *y*-direction in Fig. 1a, and to the end surfaces in both *x*-and *y*-directions in Fig. 1b. The depth of the MM simulation box in *y*-direction is 30 Å and the substrate thickness is 15 Å, larger than the cut-off radius in calculating von der Waals force. The width of the graphene portion demarcated by the dash lines and that of the underlying substrate in *x*-direction, and the nanowire diameters are varied to study their effects on the graphene morphology. The C-C bonding energy in the graphene is described by the second generation Brenner potential[24]. The interaction energy between the graphene and the nanowires and that between the graphene and the substrate are computed by the sum of the van der Waals forces between all C-Si and C-O atomic pairs in the system. These two types of van der Waals forces are described by two Lennard-Jones (LJ) pair potentials, respectively, both of which take the general form of $V(r) = 4\varepsilon(\sigma^{12}/r^{12} - \sigma^6/r^6)$, where $\sqrt[6]{2}\sigma$ is the equilibrium distance of the atomic pair and $\varepsilon$ is the bonding energy at the equilibrium distance. Parameters in the C-Si pair potential and those in the C-O pair potential are listed in Table 1. To reduce the computation cost, the bonding energy in the Si nanowire and the $SiO_2$ substrate, and the non-bonding Si-$SiO_2$ interaction are assumed to be constant, thus not considered in the energy minimization. This assumption is justified given the rigidity of Si and $SiO_2$ solids compared with the out-of-plane flexibility of a graphene monolayer.

In each MM simulation case, the graphene is prescribed with an initial morphology that partially conforms to the envelop defined by the nanowire and the substrate surface (e.g., in Fig. 1). The carbon atoms in the graphene then adjust their spatial positions to minimize the system energy, and eventually define the equilibrium morphology of the graphene. The total energy of the system is minimized using the limited-memory Broyden-Fletcher-Goldfarb-Shanno (BFGS) algorithm[25], until the total net force is less than $10^{-6}$ eV/ Å. The MM simulations are carried out by running a code in a high performance computer cluster.

## 3. RESULTS AND DISCUSSION

### 3.1. Morphological interaction between a graphene nanoribbon and a Si nanowire on a SiO$_2$ substrate

We consider the interaction between a graphene nanoribbon of finite width $W$ and a Si nanowire of diameter $d$ on a SiO$_2$ substrate. Such a structure can be fabricated by transfer printing a mechanically exfoliated graphene nanoribbon onto a SiO$_2$-supported Si nanowire[26], with the length direction of the graphene nanoribbon parallel to the axial direction of the nanowire. As will be shown in this section, the regulated morphology of the graphene nanoribbon depends strongly on its width $W$ and the nanowire diameter $d$.

If $W$ is smaller than or comparable to $d$, the morphology of such a narrow graphene nanoribbon is mainly determined by the interaction between the graphene and the nanowire, and that between the graphene and the substrate becomes negligible. Depending on the relative value of $W$ and $d$, the narrow graphene nanoribbon can have two different morphologies, as illustrated in Fig. 2. For example, on a Si nanowire of $d = 4$ nm, a graphene nanoribbon of $W = 6$ nm remains nearly flat (Fig. 2a). By contrast, on Si nanowire of $d = 10$ nm, a graphene nanoribbon of $W = 12$ nm fully conforms to the surface of the Si nanowire (Fig. 2b). These two different morphologies of the graphene nanoribbon can be explained as follows. The strain energy density of the graphene due to out-of-plane bending approximately scales with the square of the local curvature of the graphene. Therefore, the graphene strain energy due to conforming to the Si nanowire surface is roughly proportional to $1/d^2$. If the Si nanowire is too thin, the significant increase of the graphene strain energy can overbalance the decrease of the graphene-nanowire interaction energy due to graphene conforming to the nanowire. As a result, the graphene nanoribbon remains nearly flat on the Si nanowire. On the other hand, if the Si nanowire is sufficiently thick, the decrease of the interaction energy outweighs the increase of the graphene strain energy. Consequently, the graphene nanoribbon conforms to the nanowire surface. Also note when the graphene nanoribbon remains nearly flat on the Si nanowire, the two free edges in $y$-direction form ripples due to the edge stress[27] in the graphene nanoribbon.

If $W$ is much larger than $d$, the equilibrium morphology of the graphene nanoribbon takes the form as shown in Fig. 3a. The graphene portion far away from the Si nanowire conforms to the flat surface of the $SiO_2$ substrate while the middle portion of the graphene partially wraps around the Si nanowire. The geometry of the graphene-nanowire-substrate system at the equilibrium can be characterized by three parameters: the width of the corrugated portion of the graphene nanoribbon $L$, the width of the graphene nanoribbon $W$, and nanowire diameter $d$ (Figure 3a). Figure 3b plots $L/d$ as a function of $d$ for various widths of graphene nanoribbon $W/d = 6.8, 7.0, 7.2$ and $7.4$, respectively. When the graphene nanoribbon is sufficiently wide (e.g., much larger than $d$), $L$ is roughly independent of $W$, as evident with the small variation among the results for the four different values of $W$. As shown in Fig. 3b, $L/d$ decreases as $d$ increases, and then approaches to a plateau of about 2.2 when $d$ exceeds 7 nm. Such a trend can be explained by the similar argument aforementioned. If the Si nanowire is too thin, the graphene nanoribbon can only barely wrap around the nanowire, given the significant constraint of possible strain energy increase in the graphene. The corrugated portion of the graphene ribbon gradually transits to the flat portion on the substrate surface, resulting in a relatively large $L/d$ (e.g., 3.7 for $d = 2$ nm). If the Si nanowire is thick enough, the graphene nanoribbon can wrap more of the nanowire surface, leading to a higher slope of the graphene sagging down toward the substrate surface, and thus a smaller $L/d$. As the nanowire diameter is greater than ~7 nm, $L/d$ tends to a constant of

~2.2. In other words, the morphology of the corrugated portion of the graphene nanoribbon intercalated by a sufficiently thick nanowire is self-similar.

Figure 4 further plots the Brenner potential energy of the carbon atoms in the graphene nanoribbon at the equilibrium state (e.g., Fig. 3a), which depicts the strain energy profile of the corrugated graphene. As indicated by the color shades, high strain energy states in the graphene nanoribbon occur at the regions near the top of the nanowire and where the graphene becomes flat on the substrate. At such regions the graphene nanoribbon bends the most. The inset of Fig. 4 further shows the average C-C bond lengths in *x*-direction at three locations, as indicated by three marked lines A (where graphene remains flat on the substrate), B (in the middle of the intercalated portion of the graphene) and C (the crest of the corrugated graphene portion). Given the equilibrium C-C bond length of 1.42 Å, it is evident that the C-C bonds are stretched at B in *x*-direction and un-stretched at A and C. As shown in Fig. 4, comparing with the much higher energy of the carbon atoms at C due to bending, the Brenner potential energy of the carbon atoms at B is at a comparable level of that of carbon atoms at A, even though the C-C bonds at B are stretched but those at C not . Therefore, it can be estimated that the strain energy of the corrugated graphene nanoribbon is mainly due to bending, rather than stretching.

**3.2. Morphologic interaction between a blanket graphene flake and an array of Si nanowires evenly patterned in parallel on a SiO$_2$ substrate**

When a blanket graphene flake is intercalated by an array of Si nanowires evenly patterned in parallel on a SiO$_2$ substrate, the nanowire spacing $W$ comes into play in determining the regulated morphology of the graphene flake. Emerging from the simulations are two types of morphologies of graphene at equilibrium, depending on $W$ and $d$, as shown in Figure 5.

If the nanowires are widely spaced (e.g., $W>>d$), the graphene tends to conform to the envelop of each individual nanowire (Figure 5a), sags down and adheres to the substrate in between neighboring nanowires. The corrugated portion of the graphene is of a width of $L$ and an amplitude of $A_g$ ( $\approx d$ in this case). Figure 6a further plots $L/d$ as a function of $d$ for various values of $W$. For a given $W$, $L/d$ increases as $d$ increases in a roughly linear manner. When compared with the case of graphene nanoribbon intercalated by a single nanowire on a substrate (e.g., Fig. 3b), the width of the corrugated portion of the graphene intercalated by patterned nanowires on a substrate is much larger. This can be explained by the constraint of the portion of the graphene sagged in between neighboring nanowires and adhered to the substrate. Therefore, the graphene cannot slide easily on the substrate to conform to the envelop of each individual nanowire closely. As a result, the corrugated portion of the graphene is under modest stretch in $x$-direction. By contrast, the graphene nanoribbon intercalated by a single nanowire is shown to be able to slide on the substrate surface to conform more to the nanowire surface. As a result, the stretch in the graphene in $x$-direction can be nearly fully relaxed (e.g., Fig. 4), leading to a much smaller value of $L/d$. Figure 6a also indicates the increase of $L/d$ as $W$ decreases, for a given value of $d$. To further clarify this trend, Fig. 6b plots $L/d$ as a function of $W/d$ for various values of $d$, which indicates a roughly linear dependence between $L$ and $W$, as well as a weak dependence on $d$. This can be explained that, when nanowire spacing is larger, the corrugation-induced graphene stretching is accommodated by a longer graphene segment, leading to smaller strain energy of the graphene. Therefore the graphene conform more to the nanowires. The limiting case of infinitely large nanowire spacing corresponds to that of graphene intercalated by a single nanowire (e.g., Fig. 3).

If the spacing between the patterned nanowires is not sufficiently large, the graphene flake remains nearly flat, just slightly conforming to the envelop of the nanowires with a negligible amplitude $A_g$ (Fig. 5b), a morphology of graphene distinct from that regulated by widely distributed nanowires on a substrate (i.e., Fig. 5a). For a given nanowire diameter $d$, there is a sharp transition between these two distinct morphologies as the nanowire spacing reaches a critical value $W_{cr}$. This result agrees with the snap-through instability of the graphene morphology regulated by patterned nanowires on a substrate predicted by our previous continuum model results[20]. Such a snap-through instability of the graphene morphology results from the double-well energy profile of the system[19]. For the bonding parameters given in Table 1, $W_{cr}/d$ ranges from 12.3 to 12.8, and is approximately independent of $d$. In practice, it is possible to have chemical bonding or pinnings between the graphene and the substrate, the nanowire surface can also be functionalized to facilitate chemical bonding with the graphene, both of which lead to an enhanced interfacial bonding energy of the graphene. Based on the energetic interplay as described in Section 2, it is expected that the resulting graphene morphology can also be tuned by the interfacial bonding energy.

To investigate the effect of graphene-substrate interfacial bonding energy on the graphene morphology, a tuning factor $\lambda$ is used to vary the bonding energy in the LJ pair potential describing the graphene-substrate van der Waals interaction. A tuning factor $\lambda>1$ denotes a graphene-substrate interaction energy stronger than that described in Table 1. Here, the graphene-nanowire interaction energy remains the same. Figure 7 plots the normalized amplitude of graphene corrugation $A_g/d$ as a function of tuning factor $\lambda$ for $d = 4$ nm, 4.2 nm and 4.4 nm, respectively. Here, $W$=46.9 nm. For a given nanowire diameter and spacing, there exists a critical graphene-substrate interaction energy (i.e., a critical tuning factor $\lambda_c$), weaker than which the graphene only slightly conform to the envelop of the nanowires (e.g., $A_g/d \ll 1$), and stronger than which the graphene can sag in between the nanowires and adhere to the substrate (e.g., $A_g/d \approx 1$). The sharp transition between these two distinct morphologies at the critical graphene-substrate interfacial energy reveals the similar snap-through instability of the graphene morphology aforementioned. Previous studies based on continuum model[20] suggest that $W/d \propto \lambda_c^{1/4}$, or $W\lambda_c^{1/4}/d$ is a constant. The results shown in Fig. 7 give $W\lambda_c^{1/4}/d = 12.40$, 12.25, and 12.26 for $d = 4$ nm, 4.2 nm, and 4.4 nm, respectively. In this sense, the molecular mechanics simulation results agree well with the continuum model prediction.

## 4. CONCLUDING REMARKS

Using molecular mechanics simulations, we determine the morphologic interaction between a graphene nanoribbon and a Si nanowire on a $SiO_2$ substrate, and that between a blanket graphene flake and an array of Si nanowires evenly patterned in parallel on a $SiO_2$ substrate. The regulated graphene morphology is shown to be determined by the interplay between the graphene strain energy due to bending and stretching and the graphene-substrate and graphene-nanowire interaction energies. Specifically, the graphene morphology can be tuned by both geometric parameters (e.g., graphene nanoribbon width, nanowire diameter, and nanowire spacing) and material properties (e.g., graphene-nanowire and graphene-substrate bonding strength). Various simulations are conducted to quantify the relation between the graphene morphology and these parameters and properties. The present study assumes negligible deformation of the substrate under the morphologic interaction, which is justified by the rigidity of $SiO_2$. Recent experiments demonstrate the snap-through instability of the morphology of graphene on an elastomeric substrate[15]. Given the ultralow stiffness of such substrate materials (e.g., 1s~10s MPa), the elastomeric substrate may deform to some extent under the morphologic interaction with the graphene. To precisely predict the regulated morphology of the graphene in such a material system, the elastic energy of the substrate needs to be considered in the energetic interplay, which is addictive to the present model and will be reported elsewhere. The molecular mechanics simulation results agree well with our previous continuum mechanics modeling results, and furthermore, provide atomistic scale information of each carbon atoms in the graphene. Such information can be consequently used to determine the effect of the morphologic interaction and the mechanical deformation on the electronic properties of the graphene (e.g., bandgap). Though much remains to be done to achieve fine tuning of graphene's electronic properties via morphologic interaction, the present study demonstrates the beginning steps toward this promising approach that could potentially enable new graphene-based device applications.

**ACKNOWLEDGEMENTS:**

This work is supported by the Minta-Martin Foundation, a UMD General Research Board summer research award to T. L., and National Science Foundation (Grant No. 0928278). Z.Z. also thanks the support of the A. J. Clark Fellowship and the UMD Future Faculty Fellowship.

Table 1. LJ potential parameters used in molecular mechanics simulations[28]

|      | ε (eV)  | σ (Å) |
|------|---------|-------|
| C-Si | 0.00213 | 1.506 |
| C-O  | 0.00499 | 2.256 |

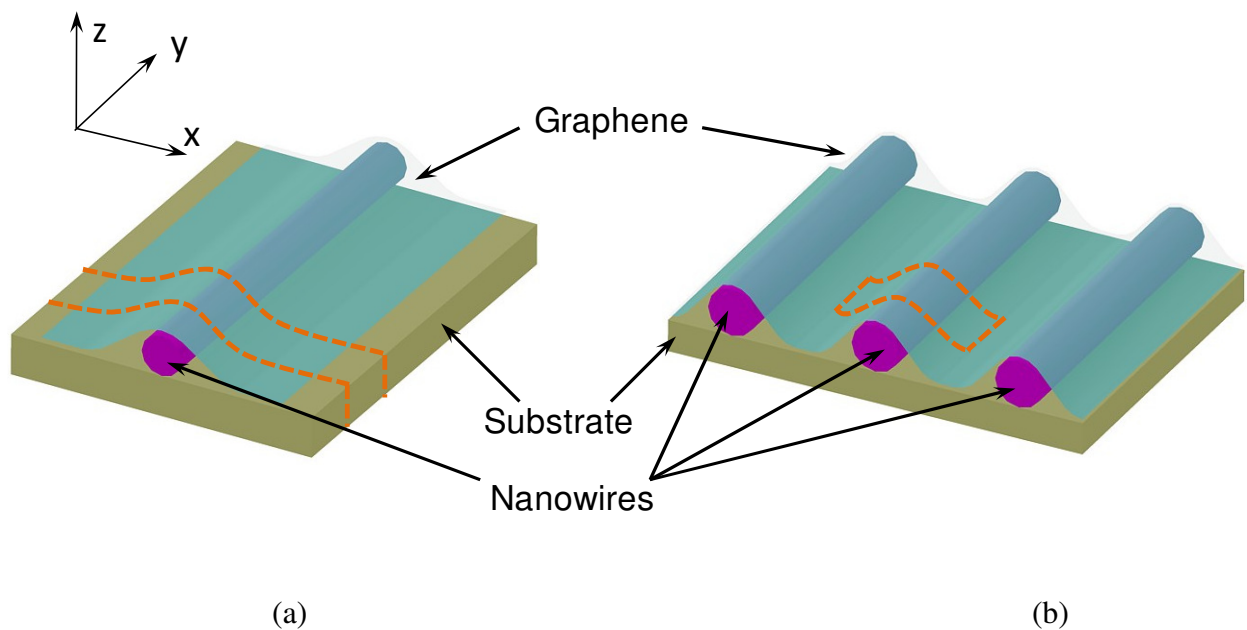

(a)                                                          (b)

Figure 1. Schematics of two simulation cases. (a) A graphene nanoribbon intercalated by a Si nanowire on a $SiO_2$ substrate; (b) A blanket graphene flake intercalated by an array of Si nanowires evenly patterned in parallel on a $SiO_2$ substrate. The dash lines delineate the portion of graphene and the underlying nanowire and substrate simulated by molecular mechanics in each case.

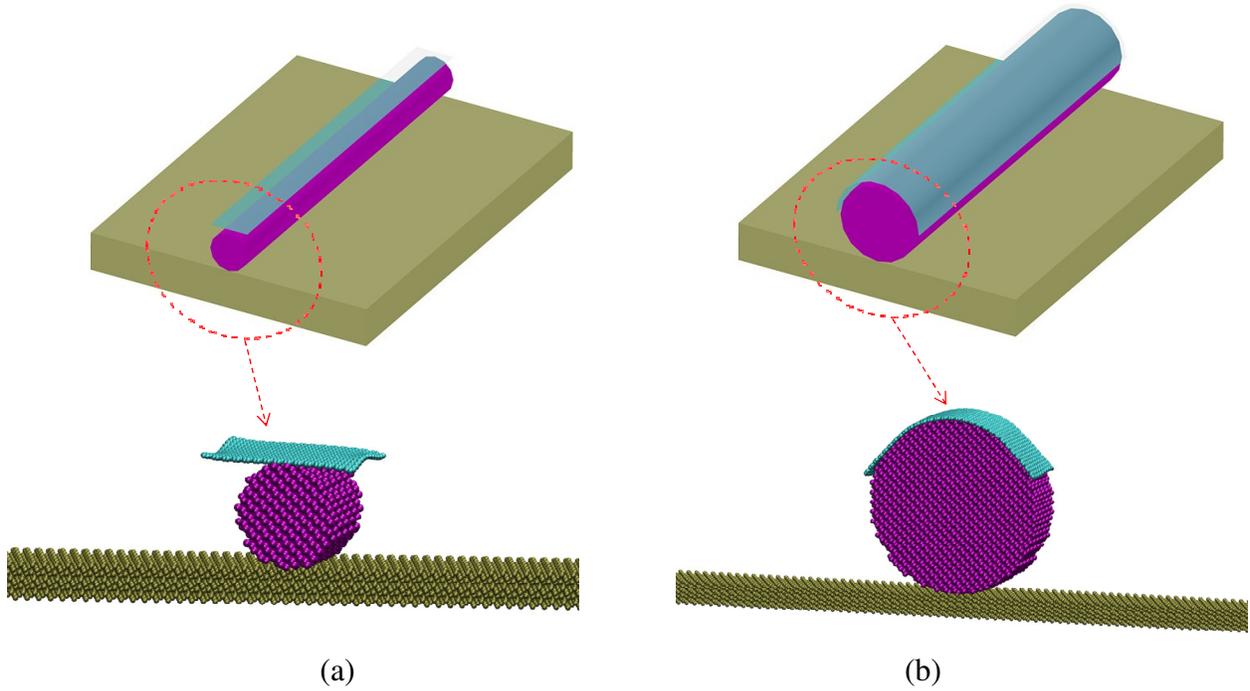

(a) (b)

Figure 2. (a) On a nanowire of diameter of 4 nm, a narrow graphene nanoribbon of width of 6 nm remains nearly flat, with slight ripples along two long edges. (b) On a nanowire of diameter of 10 nm, a narrow graphene nanoribbon of width of 12 nm can fully conform to the nanowire surface.

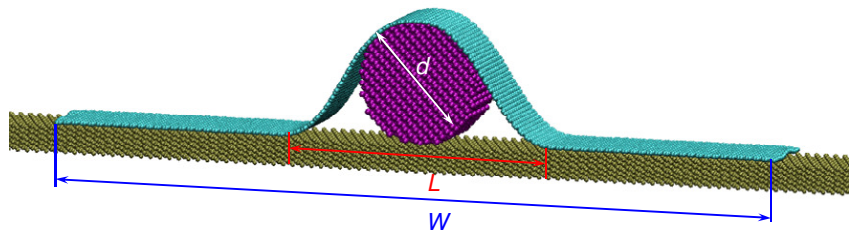

(a)

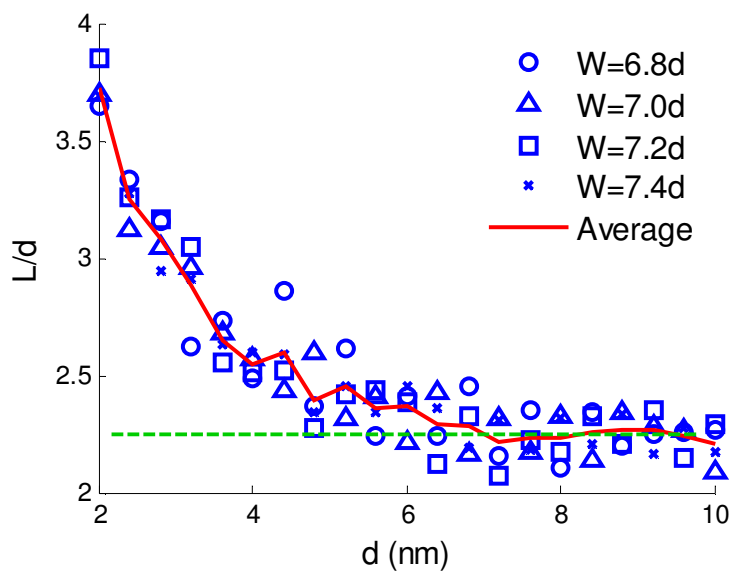

(b)

Figure 3. (a) Molecular mechanics simulation result of the morphology of a wide graphene nanoribbon intercalated by a Si nanowire on a $SiO_2$ substrate. Here $d = 6$ nm and $W = 40$ nm. (b) Normalized width of the corrugated portion of the graphene $L/d$ as a function of $d$ for various widths of the graphene nanoribbon $W = 6.8d$, $7.0d$, $7.2d$ and $7.4d$, respectively. The solid line plots the average of the four data sets. The dash line shows the plateau value of $L/d$ when $d$ is sufficiently large.

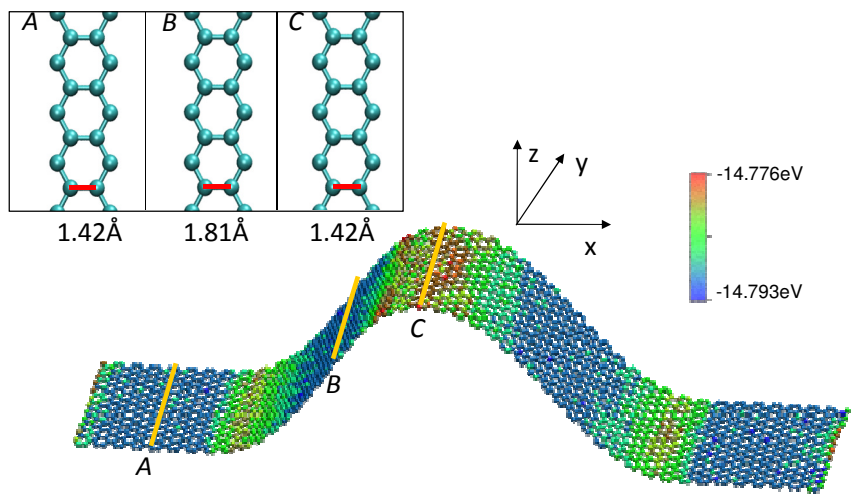

Figure 4. The distribution of Brenner potential energy of the carbon atoms in a graphene nanoribbon intercalated by a Si nanowire (not shown) on a $SiO_2$ substrate (not shown). The inset shows the C-C bond lengths in *x*-direction at three cross-sections marked by the solid lines on graphene, indicating the graphene is under tension near location B. Here $d = 4$ nm and $W = 20$ nm.

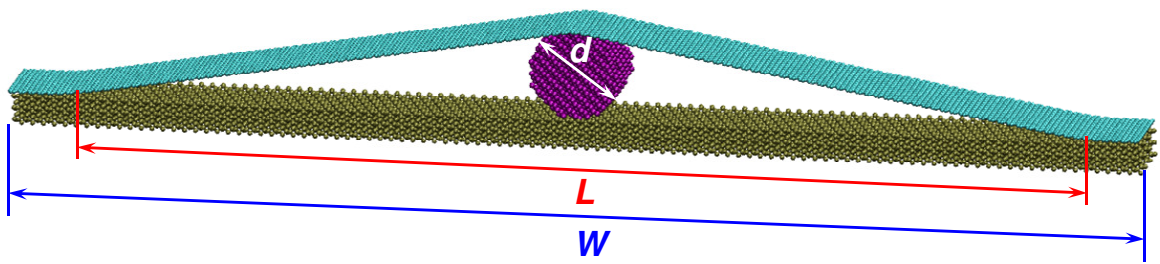

(a)

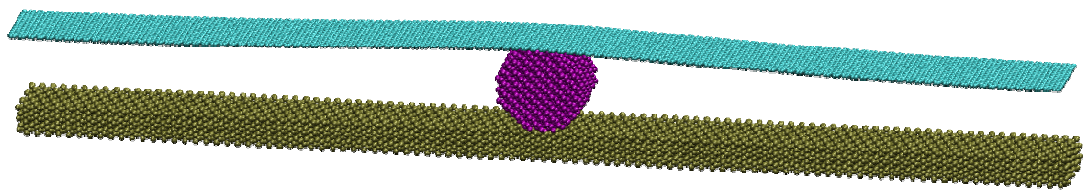

(b)

Figure 5. Molecular simulation results of the morphology of a blanket graphene flake intercalated by Si nanowires evenly patterned in parallel on a $SiO_2$ substrate. (a) When the Si nanowires are widely spaced (e.g., $W$ is large), graphene sags in between neighboring nanowires and adhere to the substrate surface. The width of the corrugated portion of the graphene is denoted by $L$. (b) If the nanowire spacing is small, graphene remains nearly flat, just slightly conform to the envelop of the nanowires. Here $d = 4$ nm and $W = 48$ nm in (a) and 46 nm in (b). the sharp transition between (a) and (b) as $W$ varies indicates a snap-through instability of the graphene morphology.

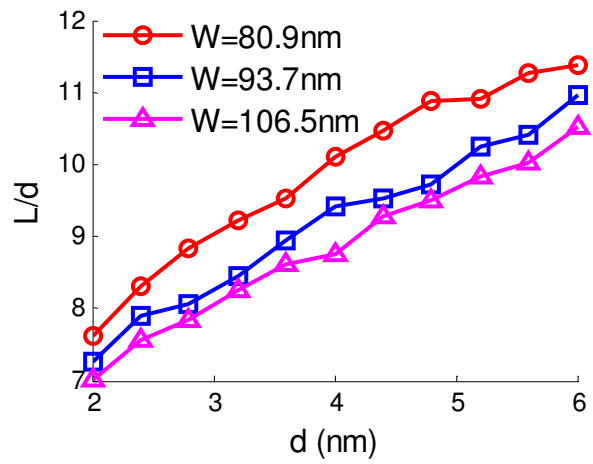

(a)

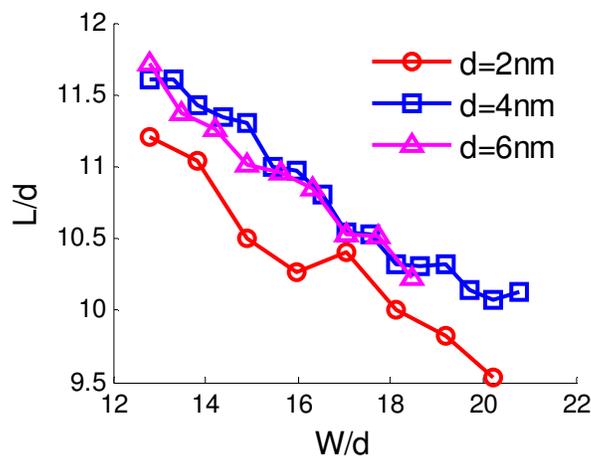

(b)

Figure 6. (a) *L/d* as a function of *d* for various values of *W*. (b) *L/d* as a function of *W/d* for various values of *d*.

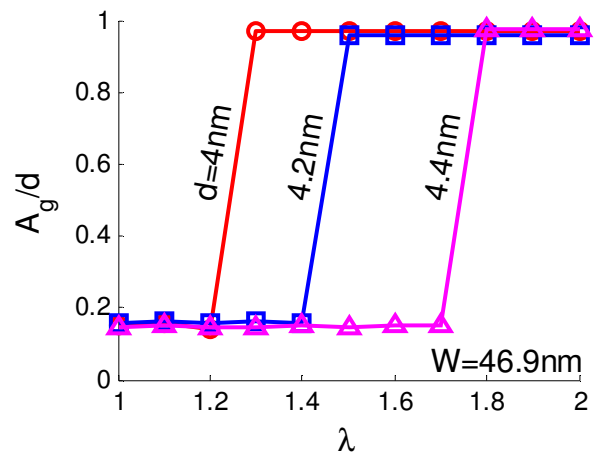

Figure 7. $A_g/d$ as a function of tuning factor $\lambda$ for various values of $d$. Here, $W$=46.9 nm. The sudden jumps of $A_g/d$ at a critical value of $\lambda$ indicate the snap-through instability of the graphene morphology shown in Fig. 5.